\documentclass[12pt, a4paper]{article}
\usepackage[cp1251]{inputenc}
\usepackage[russian]{babel}
\usepackage{indentfirst}
\usepackage{misccorr}
\usepackage[dvips]{graphicx}
\usepackage{amsmath, latexsym, amssymb, bm, array, graphics, amsfonts, amsthm, eucal}
\makeatletter
\renewcommand{\@biblabel}[1]{#1.\hfill}
\newcommand{\Res}{\mathop{\rm Res\,}}
\newcommand{\const}{\mathop{\rm const\, }}

{

\makeatother

\begin{document}
\large
\newcommand{\mc}[1]{\mathcal{#1}}
\newcommand{\E}{\mc{E}}
\renewcommand{\refname}{\begin{center} \center \rm СПИСОК ЛИТЕРАТУРЫ\end{center}}

\centerline{\bf A. V. Latyshev, S. Suleimanova}

\begin{center}
{\bf  The analytical solution of the problem on plasma oscillations in half-space
with diffusion boundary conditions}
\end{center}

\begin{center}
{\it Faculty of Physics and Mathematics,\\ Moscow State Regional
University, 105005,\\ Moscow, Radio str., 10-A}
\end{center}\medskip

\noindent The boundary problem about behaviour (oscillations) of the electronic plasmas
with arbitrary degree of degeneration of electronic gas in half-space with
diffusion boundary conditions is analytically solved.  The kinetic equation of
Vlasov---Boltzmann with integral of collisions of type BGK (Bhatnagar, Gross, Krook) and
Maxwell equation for electric field are applied.
Distribution  function for electrons and electric field in plasma in the form of
expansion under eigen  solutions of the initial system of equations  are received.
Coefficients of these expansions are found by means of the boundary conditions. \bigskip

\bigskip

\noindent {\it УДК 519.634}
\begin{center}
{\bf АНАЛИТИЧЕСКОЕ РЕШЕНИЕ ЗАДАЧИ О КОЛЕБАНИЯХ ПЛАЗМЫ  В ПОЛУПРОСТРАНСТВЕ
С ДИФФУЗНЫМИ ГРАНИЧНЫМИ УСЛОВИЯМИ}
\end{center}

\centerline{\bf \copyright \;2016 г. \quad А. В. Латышев, С. Сулейманова}

\begin{center}
  {(\it 105005, Москва, ул. Радио, 10а, МГОУ)\\
e-mail: avlatyshev@mail.ru, sevda-s@yandex.ru\\}
\end{center}

\noindent Аналитически решена граничная задача о поведении (колебаниях) электронной
плазмы с произвольной степенью вырождения электронного газа в полупространстве с
диффузными граничными условиями. Применяются кинетическое уравнение
Власова---Больцмана с интегралом столкновений типа БГК (Бхатнагар, Гросс, Крук) и
уравнение Максвелла для электрического поля. Функция распределения электронов и
электрическое поле внутри плазмы получены в виде разложений по собственным решениям
исходной системы уравнений. Коэффициенты этих разложений найдены с помощью граничных
условий. \bigskip

\noindent {{\bf Ключевые слова}: уравнение Власова---Больцмана, уравнение Максвелла,
частота столкновений, электромагнитное поле, моды Друде, Дебая, Ван Кампена,
дисперсионная функция, краевая задача Римана.}\bigskip

\begin{center}
{ВВЕДЕНИЕ}
\end{center}

Настоящая работа является продолжением ряда работ, посвященных проблеме поведения
вырожденной и невырожденной (максвелловской) плазмы в полупространстве, на границе
которого задано внешнее продольное электрическое поле (см., например,
\cite{Lat2001}--\cite{Lat2007}). В \cite{Lat2001} и \cite{Lat2006b} рассмотрен
случай вырожденной плазмы, а в \cite{Lat2006a} и \cite{Lat2007} рассмотрен случай
невырожденной максвелловской плазмы. Случай плазмы с произвольной степенью
вырождения электронного газа до сих пор не рассматривался. Данная работа
представляет собой попытку устранить этот пробел. В настоящей работе рассматривается
общий случай плазмы с произвольной степенью вырождения электронного газа. Так что в
этом смысле данная работа является завершающей.

Везде ниже в статье плазму будем считать электронной. Согласно \cite{Landau10}, плазма
называется электронной, если в диэлектрической поляризации плазмы участвуют только
электроны, а движение ионов несущественно и им можно пренебречь. Понятие "плазма"\,
появилось в работах Тонкса и Лэнгмюра \cite{Tonks}. Колебания плазмы изучались в
работах Власова (см., например, \cite{Vlasov}). Как граничная задача математической
физики задача о колебаниях плазмы с зеркальным условием отражения электронов корректно
поставлена в работе \cite{Landau46}. Работы \cite{Gohfeld1} и \cite{Gohfeld2}
посвящены дальнейшему изучению поведения плазмы -- затуханию электромагнитных волн в
плазме и их аномальному проникновению внутрь плазмы.

В настоящей работе получено аналитическое решение задачи о поведении плазмы с
произвольной степенью вырождения электронного газа в полупространстве во внешнем
переменном продольном электрическом поле. Используются кинетическое уравнение
Власова---Больцмана с интегралом столкновений типа БГК (Бхатнагар, Гросс и Крук)
\cite{BGK} и уравнение Максвелла для электрического поля.

Выяснена структура экранированного электрического поля. Оказалось, что существует
область значений параметров задачи, в которой отсутствует мода Дебая.

Показано, что, во-первых, мода Друде, описывающая объемную проводимость, существует
при всех значениях параметров задачи; во-вторых, мода Дебая, описывающая экранировку
электрического поля, существует при частотах колебания внешнего поля, меньших
некоторой критической частоты, находящейся вблизи плазменного резонанса; и, наконец,
что моды Ван Кампена \cite{VanKampen}, представляющие собой смесь (хаотизацию)
собственных решений уравнения Власова---Больцмана, 
также существуют при всех значениях параметров задачи. Оказалось, что моды Ван
Кампена отвечают непрерывному спектру характеристического уравнения, мода Дебая
отвечает дискретному спектру задачи, мода Друде --- спектру, присоединенному к
непрерывному. Показано, что аналитическое решение граничной задачи может быть
представлено в виде разложения по собственным решениям, отвечающим перечисленным
выше спектрам.

\begin{center}
{1. ПОСТАНОВКА ЗАДАЧИ И ОСНОВНЫЕ УРАВНЕНИЯ}
\end{center}

Пусть невырожденная плазма Ферми---Дирака занимает полупространство $x>0$.
Будем использовать $\tau$--модельное уравнение Власова---Больцмана:
$$
\frac{\partial f}{\partial t}+\mathbf v \frac{\partial f}{\partial \mathbf r}+
e\left(\mathbf E+\frac{1}{c}[\mathbf v, \mathbf H]\right)
\frac{\partial f}{\partial \mathbf p}=\nu(f_{eq}-f)
\eqno{(1.1)}
$$
и уравнение Максвелла для электрического поля
$$
\mathrm{div} \mathbf E = 4\pi e\int (f - f_0)d\Omega_F, \quad
d\Omega_F=\frac{(2s+1)d^3p}{(2\pi\hbar)^3}.
\eqno {(1.2)}
$$

Здесь $f_{eq}$ -- локально--равновесная функция распределения Фер\-ми---Дирака,
$$
f_{eq}({\bf r},v,t)=\left\{1+\mathrm {exp}\frac{\mathcal E-\mu ({\bf r},t)}{kT}\right\}^{-1},
$$
$f_0 = f_{FD}$ -- невозмущенная функция распределения Ферми---Дирака,
$$
f_0(v,\mu) = f_{FD}(v, \mu) =\left \{1+\mathrm {exp}\frac{\mathcal E-\mu}{kT}\right\}^{-1},
$$
{\bfseries p} = {\itshape m}{\bfseries v} -- импульс электрона,
${\mathcal E}$=${mv^2}/{2}$ -- кинетическая энергия электрона, $\mu=\const$ и $\mu({\bf r},t)$ --
соответственно невозмущенный и возмущенный химический потенциал, $e$ и $m$ -- заряд и
масса электрона, $\rho$ -- плотность заряда, $\hbar$ -- постоянная Планка, -- эффективная
частота рассеяния электронов, {\itshape s} -- спин частиц, для электрона
{\itshape s}=1/2, {\itshape k} -- постоянная Больцмана, {\itshape T} -- температура плазмы,
которая считается постоянной в данной задаче, {\bfseries E}({\itshape {\bf r},t}) и
{\bfseries H}({\itshape {\bf r},t}) -- электрическое и магнитное поля внутри плазмы.

Пусть внешнее электрическое поле вне плазмы перпендикулярно границе плазмы
и меняется по закону
$$
\mathbf E_{\rm ext}(t)=E_0e^{-i\omega t}(1,0,0).
$$
Соответствующее электрическое поле внутри плазмы будем обозначать через
$$
E(x,t)=E(x)e^{-i\omega t}.
$$
Так как внешнее поле имеет одну {\itshape x}--компоненту, то функция распределения $f$
имеет вид $f = f(x,v_x,t)$, $v_x$ -- проекция скорости электронов на ось $x$.
Внешнее электрическое поле вызывает изменение химического потенциала
$$
\mu (x,t) = \mu + \delta\mu(x,t).
$$

Здесь $\mu=\const$ -- значение химического потенциала, отвечающее отсутствию внешнего
электрического поля на границе плазмы.

Введем безразмерный импульс (скорость) электронов ${\mathbf P}$=${\mathbf p}/{p_T}=
{\mathbf v}/{v_T}$, $v_T$ -- тепловая скорость электронов, $v_T$=$\sqrt{2kT/m}$
и безразмерный (приведенный) химический потенциал $\alpha={\mu}/{kT}$.
Для приведенного химического потенциала предыдущее равенство имеет вид
$$
\alpha(x,t) = \alpha + \delta \alpha (x,t).
$$

Будем считать, что величина $\delta\alpha (x,t)$ -- возмущение приведенного химического
потенциала является малым параметром, т.е.
$$
|\delta\alpha (x,t)|\ll1.
$$

Физически это
неравенство означает, что возмущение химического потенциала много меньше тепловой
энергии электронов:
$$
|\delta\mu(x,t)|\ll\mathcal E_T,\; \mathcal E_T={mv^2_T}/{2}.
$$
 Будем действовать методом последовательных приближений, считая, что $|\delta\alpha(x,t)|\ll1$.

Линеаризацию уравнений (1.1) и (1.2) проведем относительно абсолютной функции
распределения Ферми---Дирака
$$
f_0(P,\alpha)=f_{FD}(P,\alpha)=\frac{1}{1+e^{P^2-\alpha}}.
$$

Линеаризуя локально--равновесную функцию распределения, получаем, что
$$
f_{eq}(x,P,t)=f_0(P,\alpha)+g(P,\alpha)\delta \alpha(x)e^{-i\omega t},
$$
где
$$
f_0(P,\alpha)=f_{FD}(P,\alpha)=\dfrac{1}{1+e^{P^2-\alpha}},$$$$
g(P,\alpha)=e^{P^2-\alpha}(1+e^{P^2-\alpha})^{-2}.
$$

Заметим, что в линейном приближении
$$
e\left(\mathbf E+\frac{1}{c}[\mathbf v,\mathbf H]\right)
\frac{\partial f}{\partial\mathbf p}=\dfrac{e}{p_T}{\mathbf E}
\frac{\partial f_0}{\partial\mathbf P}=$$$$=\dfrac{e}{p_T}E(x)e^{-i\omega t}
\frac{\partial f_0}{\partial P_x}=
-\dfrac{2eP_x}{p_T}E(x)e^{-i\omega t}g(P,\alpha).
$$

Линеаризуем функцию распределения электронов:
$$
f(x,\mathbf P,t)=f_0(P,\alpha)+g(P,\alpha)h(x,P_x)e^{-i\omega t}.
\eqno {(1.3)}
$$
Здесь $h(x,P_x)$ - новая неизвестная функция.

Вместо уравнений (1.1) и (1.2) с помощью (1.3) и предыдущих соотношений получаем:

$$
v_TP_x\frac{\partial h}{\partial x}+(\nu-i\omega) h(x,P_x)=
E(x)\frac{2eP_x}{p_T}+\nu \delta\alpha(x),
\eqno {(1.4)}
$$
$$
\frac{d E(x)}{d x}=\frac{8\pi e p^3_T}{(2\pi\hbar)^3}\int h(x,P_x)g(P,\alpha)d^3P.
\eqno {(1.5)}
$$

Величина $\delta\alpha(t,x)$ определяется из закона сохранения числа частиц:
$$
\int f_{eq}d\Omega_F=\int f d\Omega_F.
\eqno {(1.6)}
$$

Из (1.6) находим, что
$$
\delta\alpha(x)=\frac{\int h(x,P_x)g(P,\alpha)d\Omega_F}{\int g(P,\alpha)d\Omega_F}.
$$

Заметим, что
$$
\int g(P,\alpha)d^3P=\int_{-1}^{1}d\mu \int_{0}^{2\pi}d\chi
\int_{0}^{\infty}\frac{e^{P^2-\alpha}P^2dP}{\left(1+e^{P^2-\alpha}\right)^2}=$$$$=
4\pi\int_{0}^{\infty}\frac{e^{P^2-\alpha}P^2dP}{\left(1+e^{P^2-\alpha}\right)^2}=
4\pi g_2(\alpha),
$$
где
$$
g_2(\alpha)=\int_{0}^{\infty}\frac{e^{P^2-\alpha}P^2dP}
{\left(1+e^{P^2-\alpha}\right)^2}=\int_{0}^{\infty}g(P,\alpha)P^2dP,
$$
и
$$
g_2(\alpha)=\frac{1}{2}\int_{0}^{\infty}\frac{dP}{1+e^{P^2-\alpha}}=
\frac{1}{2}s_0(\alpha),
$$
$$
s_0(\alpha)=\int_{0}^{\infty}\frac{dP}{1+e^{P^2-\alpha}}=\int f_0(P,\alpha)dP.
$$

Подставляя это равенство в (1.6), находим:
$$
\delta\alpha(x)=\frac{1}{4\pi g_2(\alpha)}\int h(x,P_x)g(P,\alpha)d^3P=$$$$=
\dfrac{1}{2s_0(\alpha)}\int_{-\infty}^{\infty}f_0(P_x,\alpha)h(x,P_x)dP_x.
\eqno {(1.7)}
$$
При этом было использовано равенство
$$
\int_{-\infty}^{\infty}\int_{-\infty}^{\infty}g(P,\alpha)dP_ydP_z=
\pi f_0(P_x,\alpha).
$$

Положим далее $E(x)=E_0e(x)$. Обозначим
$$
k(\mu,\alpha)=\dfrac{f_0(\mu,\alpha)}{2s_0(\alpha)},
$$
где $\mu=P_x$.

Функция $k(\mu,\alpha)$ обладает следующей нормировкой
$$
\int_{-\infty}^{\infty}k(\mu,\alpha)d\mu=1.
$$

Теперь учитывая (1.7),
вместо (1.4) и (1.5) получаем следующую систему уравнений:
$$
v_T \mu\dfrac{\partial h}{\partial x}+(\nu-i\omega)h(x,\mu)=\dfrac{2eE_0}{p_T}\mu e(x)+
\nu\int_{-\infty}^{\infty}k(\mu,\alpha)h(x,\mu)d\mu,
$$
$$
E_0\dfrac{d e(x)}{d x}=\dfrac{16\pi^2ep_T^3s_0(\alpha)}{(2\pi\hbar)^3}\int_{-\infty}^{\infty}
k(\mu,\alpha)h(x,\mu)d\mu.
$$

В этих уравнениях  перейдем к безразмерным величинам и функциям:
$$
t_1=\nu t=\frac{t}{\tau},\qquad
x_1=\frac{\nu x}{v_T}=\frac{x}{\tau v_T}=\frac{x}{l},\qquad
l=v_T\tau,
$$
$$
e(x)=\frac{E(x)}{E_0},\qquad
H(x,\mu)=\frac{\nu kT}{eE_0v_T}h(x,\mu)=
\frac{\nu p_T}{2eE_0}h(x,\mu).
$$

В результате перехода к безразмерным параметрам и функциям получаем следующую систему
уравнений:
$$
\mu\dfrac{\partial H}{\partial x_1}+w_0H(x,\mu)=\mu e(x_1)+
\int_{-\infty}^{\infty}k(\mu',\alpha)H(x,\mu')d\mu',
\eqno{(1.8)}
$$
$$
\dfrac{d e(x_1)}{d x_1}=
\varkappa^2(\alpha)\int_{-\infty}^{\infty}k(\mu',\alpha)H(x_1,\mu')d\mu'.
\eqno{(1.9)}
$$

В уравнениях (1.8) и (1.9)
$$
\varkappa^2(\alpha)=\dfrac{32\pi^2e^2p_T^3s_0(\alpha)}{(2\pi\hbar)^3m\nu^2},
$$
$$
w_0=1-i\dfrac{\omega}{\nu}=1-i \omega \tau=1-i\dfrac{\Omega}{\varepsilon},
$$
где $\Omega=\omega/\omega_p$,
$\varepsilon=\nu/\omega_p$, $\omega_p$ -- плазменная (ленгмюровская) частота,
$
\omega_p=\sqrt{4\pi e^2 N/m}.
$
Здесь $N$ -- числовая плотность (концентрация) электронов в равновесном состоянии.

Выразим параметр $\varkappa$ через плазменную частоту.
Из определения числовой плотности вытекает, что
$$
N=\int f_0(P,\alpha)d\Omega_F=\frac{2p^3_T}{(2\pi\hbar)^3}\int\frac{d^3P}{1+e^{P^2-\alpha}}=
$$
$$=
\frac{8\pi p^3_T}{(2\pi\hbar)^3}\int_{0}^{\infty}
\frac{P^2dP}{1+e^{P^2-\alpha}}=\frac{8\pi p^3_T}{(2\pi\hbar)^3}s_2(\alpha),
$$
где
$$
s_2(\alpha)=\int_{0}^{\infty}\frac{P^2dP}{1+e^{P^2-\alpha}}=\frac{1}{2}l_0(\alpha),
$$
$$
l_0(\alpha)=\int_{0}^{\infty}\ln(1+e^{\alpha-P^2})dP.
$$

Следовательно, числовая плотность частиц плазмы и тепловое волновое число
$k_T={mv_T}/{\hbar}$ связаны соотношением
$$
N=\frac{l_0(\alpha)}{2\pi^2}k^3_T=\frac{s_2(\alpha)}{\pi^2}k^3_T,
$$
откуда
$$
k^3_T=\frac{2\pi^2}{l_0(\alpha)}N=\frac{\pi^2}{s_2(\alpha)}N.
$$

Теперь нетрудно найти, что
$$
\varkappa^2(\alpha)=\frac{\omega^2_p}{\nu^2} \cdot \frac{s_0(\alpha)}{s_2(\alpha)}=
\frac{\omega^2_p}{\nu^2} \cdot \frac{1}{r(\alpha)}=\dfrac{1}{\varepsilon^2r(\alpha)},
$$
где
$$
r(\alpha)=\dfrac{s_2(\alpha)}{s_0(\alpha)}, \qquad \varepsilon=\dfrac{\nu}{\omega_p}.
$$

Известно, что частота пламенных колебаний как правило много больше частоты столкновений
электронов в металле \cite{Landau10}. Поэтому в случае, когда $\omega\sim\omega_p$
выполняется условие $\omega_p \gg \nu$.

Рассмотрим условие диффузного отражения электронов от границы полупространства:
$$
f(x=0,{\bf v},t)=f_{eq}(x=0, v,t), \qquad v_x>0.
$$
Линеаризуя это граничное условие, получаем:
$$
H(0,P_x)=\delta \alpha(0), \qquad 0<P_x<1.
$$
Обозначим $A=\delta \alpha(0)$. Далее диффузное граничное условие будем
рассматривать в виде:
$$
H(0,\mu)=A,\qquad 0<\mu<1.
\eqno{(1.10)}
$$

Граничное условие для поля на поверхности плазмы имеет вид:
$$
e(0)=1,
\eqno {(1.11)}
$$
а вдали от поверхности поле предполагается ограниченным:
$$
e(+\infty) = e_{\infty}, \qquad |e_\infty|<+\infty.
\eqno {(1.12)}
$$

Ниже нам понадобится в качестве граничного условия условие непротекания электронов
через границу плазмы:
$$
\int v_x f(x, \mathbf v, t)d\Omega_F=0.
$$

Отсюда получаем следующее интегральное условие:
$$
\int_{-\infty}^{\infty}\mu'H(0,\mu')f_0(\mu',\alpha) d\mu' = 0.
\eqno {(1.13)}
$$

Условие (1.13) является условием непротекания электронов через границу плазмы.

\begin{center}
{2. СОБСТВЕННЫЕ ФУНКЦИИ НЕПРЕРЫВНОГО СПЕКТРА}
\end{center}

Сначала будем искать общее решение системы уравнений (1.8) и (1.9).

Разделение переменных согласно общему методу Фурье приводит к следующей подстановке:
$$
H_\eta(x,\mu)=\mathrm{exp}\left(-\frac{w_0x}{\eta}\right)\Phi(\eta,\mu), $$$$
e_\eta(x)=\mathrm{exp}\left(-\frac{w_0x}{\eta}\right)E(\eta),
\eqno {(2.1)}
$$
где $\eta$ - спектральный параметр, или параметр разделения, вообще говоря комплексный.

Подставим равенства (2.1) в уравнения (1.8) и (1.9).

Получим характеристическую систему уравнений

$$
(\eta-\mu)\Phi(\eta,\mu)=\eta\mu\frac{E(\eta)}{w_0}+\frac{\eta}{w_0}
\int_{-\infty}^{\infty}k(\mu',\alpha)\Phi(\eta,\mu')d\mu',
\eqno {(2.2)}
$$

$$
-\frac{w_0}{\eta}E(\eta)=\frac{1}{\varepsilon^2r(\alpha)}
\int_{-\infty}^{\infty}k(\mu', \alpha)\Phi(\eta,\mu')d\mu'.
\eqno{(2.3)}
$$

Обозначим:
$$
n(\eta)=\int_{-\infty}^{\infty}k(\mu',\alpha)\Phi(\eta,\mu')d\mu'.
$$

С помощью этого обозначения перепишем характеристическую систему уравнений (2.2) и (2.3)
в следующем виде:
$$
(\eta-\mu)\Phi(\eta,\mu)=\frac{E(\eta)}{w_0}\mu\eta+\frac{\eta n(\eta)}{w_0r(\alpha)},
\eqno {(2.4)}
$$
$$
-\frac{w_0}{\eta}E(\eta)=\frac{n(\eta)}{\varepsilon^2r(\alpha)}.
\eqno {(2.5)}
$$

Обозначим:
$$
\eta^2_1\equiv\eta_1(\alpha)=w_0\varepsilon^2\frac{s_2(\alpha)}{s_0(\alpha)}=
w_0\varepsilon^2r(\alpha)=\varepsilon(\varepsilon-i\Omega)r(\alpha).
$$

Из систем уравнений (2.4) и (2.5) получаем следующее уравнение:
$$
(\eta-\mu)\Phi(\eta,\mu)=\frac{E(\eta)}{w_0}(\eta\mu-\eta^2_1).
\eqno {(2.6)}
$$

При $\eta\in(-\infty,+\infty)$ ищем решение уравнения (2.6) в пространстве обобщенных
функций \cite{Zharinov}:
$$
\Phi(\eta,\mu)=\frac{E(\eta)}{w_0}(\mu\eta-\eta^2_1)P\frac{1}{\eta-\mu}+
g(\eta)\delta(\eta-\mu).
\eqno {(2.7)}
$$

В равенстве (2.7) $\eta\in(-\infty,+\infty)$, $\mu\in(-\infty,+\infty)$. Множество значений
$\eta$, заполняющих числовую прямую $-\infty<\eta<+\infty$ называют непрерывным
спектром характеристического уравнения.

В (2.7) $\delta(x)$ -- дельта-функция Дирака, символ $Px^{-1}$  означает главное
значение интеграла при интегрировании выражения $x^{-1}$, функция $g(\eta)$ играет
роль произвольной "постоянной"\, интегрирования.

Решение (2.7) уравнения (2.6) называются собственными функциями характеристического
уравнения.

Для нахождения функции $g(\eta)$ подставим (2.7)  в определение нормировочной
функции $n(\eta)$. В результате получаем, что
$$
g(\eta)=\dfrac{\eta_1^2E(\eta)\Lambda(\eta)}{\eta k(\eta,\alpha)}.
$$
Здесь введена дисперсионная функция
$$
\Lambda(z)=\Lambda(z,\Omega,\varepsilon,\alpha)=1+\frac{z}{w_0\eta_1^2}
\int_{-\infty}^{\infty}\frac{\eta^2_1-\mu' z}{\mu'-z}f_0(\mu',\alpha)d\mu'.
\eqno {(2.8)}
$$

Собственные функции (2.7) характеристического уравнения (2.6) с помощью (2.8) представим
в виде
$$
\Phi(\eta,\mu)=\frac{E(\eta)}{w_0}
\left[P\frac{\mu\eta-\eta^2_1}{\eta-\mu}-w_0\eta_1^2\frac{\Lambda(\eta)}
{\eta k(\eta,\alpha)}\delta(\eta-\mu)\right].
\eqno {(2.9)}
$$

Семейство собственных функций $\Phi(\eta,\mu)$ характеристического уравнения отвечает
непрерывному спектру. Их часто называют "моды Ван Кампена" (см. \cite{VanKampen} и
\cite{Kadomtsev}).

Собственые функции (2.9) представим в виде:
$$
\Phi(\eta,\mu)=\frac{E(\eta)}{w_0}F(\eta,\mu),
$$
где
$$
F(\eta,\mu)=P\frac{\mu\eta-\eta^2_1}{\eta-\mu}-w_0\eta_1^2
\frac{\Lambda(\eta)}{\eta k(\eta,\alpha)}\delta(\eta-\mu).
$$

Дисперсионную функцию задачи $\Lambda(z)$ можно представить следующим образом:
$$
\Lambda(z)=1-\frac{1}{w_0}-\frac{z^2-\eta^2_1}{w_0\eta^2_1}\lambda_0(z,\alpha).
$$
Здесь введена  функция
$$
\lambda_0(z,\alpha)=1+z\int_{-\infty}^{\infty}\frac{k(\mu,\alpha)d\mu}{\mu-z}.
$$

Для ее граничных значений сверху и снизу на действительной оси справедливы формулы
Сохоцкого \cite{Gahov, Lifanov}:
$$
\lambda^\pm_0(\mu,\alpha)=\lambda_0(\mu,\alpha)\pm i\pi\mu k(\mu,\alpha).
$$

С
помощью этой формулы легко вычислить граничные значения сверху и снизу на
действительной оси дисперсионной функции задачи:
$$
\Lambda^\pm(\mu)=\Lambda(\mu)\pm i\frac{\pi}{w_0\eta^2_1}\mu k(\mu,\alpha)(\eta^2_1-\mu^2),
$$
$$
\frac{\Lambda^+(\mu)+\Lambda^-(\mu)}{2}=\Lambda(\mu).
$$

\begin{center}
{3. НУЛИ ДИСПЕРСИОННОЙ ФУНКЦИИ}
\end{center}

Отыщем нули дисперсионного уравнения
$$
\frac{\Lambda(z)}{z}=0.
\eqno {(3.1)}
$$

Нетрудно видеть, что значение дисперсионной функции в бесконечно удаленной точке
равно
$$
\Lambda_{\infty}=\Lambda(\infty)=1-\frac{1}{w_0}+\frac{1}{w^2_0\varepsilon^2}.
$$
Отсюда получаем, что
$$
\Lambda_{\infty}=\frac{-i\nu\omega+\omega^2_p-\omega^2}{(\nu-i\omega)^2}\ne0
$$
при любых $\nu\ne0$, т.е. в любой столкновительной плазме.

Это значит, что точка $z_i=\infty$ является нулем дисперсионного уравнения. Можно
считать, что эта точка принадлежит спектру, присоединенному к непрерывному спектру,
составляющему открытую часть числовой оси $(-\infty,+\infty)$.

Точке $z_i=\infty$ отвечает следующее решение исходной системы уравнений (2.5) и
(2.6):
$$
H_{\infty}(x,\mu)=\frac{E_{\infty}}{w_0} \cdot \mu, \qquad e_{\infty}=E_{\infty}.
\eqno {(3.2)}
$$

Здесь $E_{\infty}$ - произвольная постоянная.

Решение (3.2) не зависит от химического потенциала. Его естественно называть модой
Друде. Оно описывает объемную проводимость плазмы металла, рассмотренную Друде (см.,
например, \cite{Ashkroft}).

По определению, дискретным спектром характеристического уравнения (2.6) называется
множество конечных комплексных нулей дисперсионного уравнения (3.1), не лежащих на
действительной оси (разрезе дисперсионной функции).

\begin{figure}
\begin{center}
\includegraphics[width=12.0cm, height=8cm]{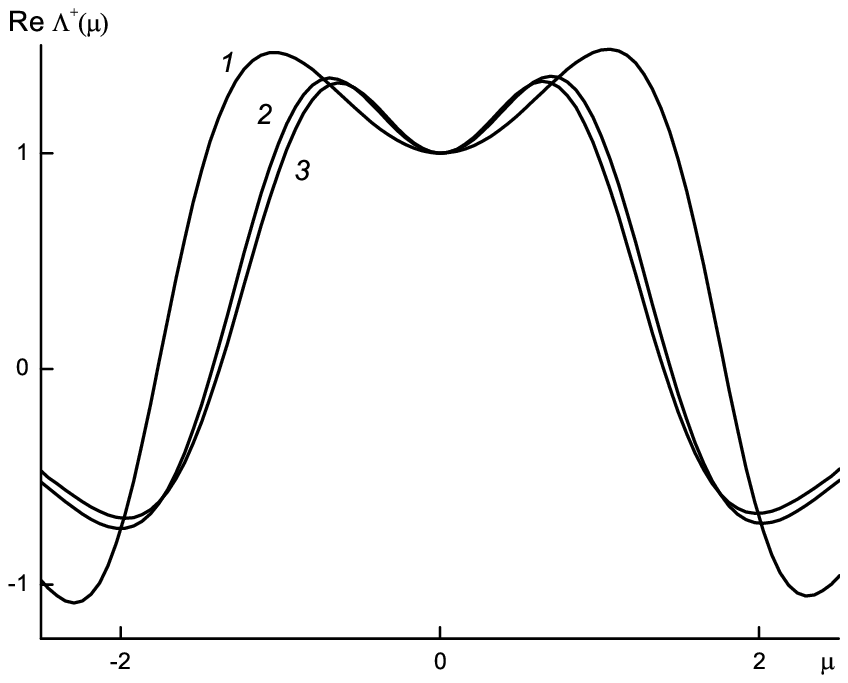}
\end{center}
\begin{center}
{Рис. 1. Действительная часть дисперсионной функции $\Lambda^{+} (\mu)$.
Кривые $1,2,3$ отвечают значениям безразмерного химического потенциала $\alpha=3,0,-1$.}
\end{center}
\begin{center}
\includegraphics[width=12.0cm, height=8cm]{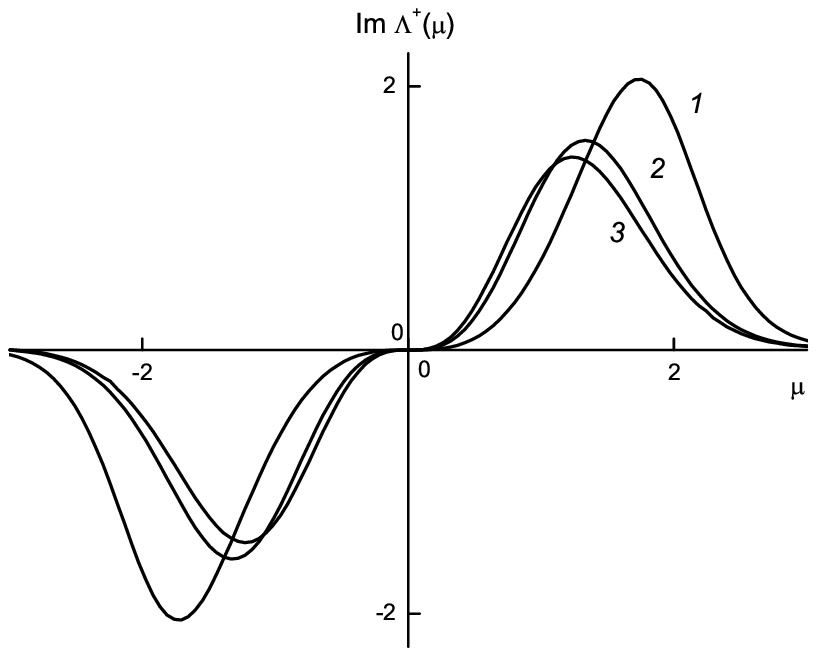}
\end{center}
\begin{center}
{Рис. 2. Мнимая часть дисперсионной функции $\Lambda^{+}(\mu)$.
Кривые $1,2,3$ отвечают значениям безразмерного химического потенциала $\alpha=3,0,-1$.}
\end{center}
\end{figure}

\clearpage

Введем обозначение:
$
\Omega=\omega/\omega_p.
$
Ясно, что
$$
\Omega\geqslant 0,\quad w_0=1-i\dfrac{\omega}{\nu}=1-i \dfrac{\Omega}{\varepsilon}=
-\dfrac{i(\Omega+i\varepsilon)}{\varepsilon}.
$$

Приступим к отысканию нулей дисперсионного уравнения.

Напишем разложение дисперсионной функции задачи в асимптотический ряд Лорана:
$$
\Lambda(z)=\Lambda_{\infty}+\frac{\Lambda_2}{z^2}+
\frac{\Lambda_4}{z^4}+ \cdot \cdot \cdot , \ \quad z\to\infty.
\eqno {(3.3)}
$$
В (3.3) введены обозначения:
$$
\Lambda_2=\frac{s_4(\alpha)-\eta^2_1s_2(\alpha)}{w_0\eta^2_1s_0(\alpha)}, \ \qquad
\Lambda_4=\frac{s_6(\alpha)-\eta^2_1s_4(\alpha)}{w_0\eta^2_1s_0(\alpha)},
\cdots.
$$
$$
s_n(\alpha)=\frac{1}{2}\int_{-\infty}^{\infty}\mu^n f_0(\mu,\alpha)d\mu=
\int_{0}^{\infty}\mu^n f_0(\mu, \alpha)d\mu.
$$

Из разложения (3.3) видно, что в окрестности бесконечно удаленной точки существует
два нуля $\pm\eta_0$ дисперсионной функции $\Lambda(z)$:
$$
\pm\eta_0\approx\sqrt{-\frac{\Lambda_2}{\Lambda_{\infty}}}.
\eqno {(3.4)}
$$
В силу четности дисперсионной функции ее нули различаются лишь знаком.
Под нулем $\eta_0$ будем понимать такое значение радикала из (3.4), что
$
\mathrm{Re}({w_0}/{\eta_0})>0.
$
Для такого нуля экспонента $\mathrm{exp}[-({w_0}/{\eta_0})x]$ является
монотонно убывающей при $x\to+\infty$.

Нулю $\eta_0$ отвечает следующее уравнение

$$
H_{\eta_{0}}(x,\mu)=\mathrm{exp}\left(-\frac{w_0}{\eta_0}x\right)\Phi(\eta_0,\mu),
$$
$$
e_{\eta_0}(x)=\mathrm{exp}\left(-\frac{w_0}{\eta_0}x\right)E_0.
$$
Здесь
$$
\Phi(\eta_0,\mu)=\frac{E_0}{w_0}\frac{\eta_0\mu-\eta^2_1}{\eta_0-\mu}.
$$

Это решение естественно назвать модой Дебая (это - плазменная мода). В
низкочастотном случае она описывает известное экранирование Дебая \cite{Ashkroft}.

Равенство (3.4) представим в явном виде:
$$
\eta_0=\eta_0(\alpha,\Omega,\varepsilon)\approx\sqrt{\frac{(\Omega+i\varepsilon)^2
[\eta^2_1s_2(\alpha)-s_4(\alpha)]}{w_0\eta^2_1s_0(\alpha)(\Omega^2-1+i\varepsilon \Omega)}}.
\eqno {(3.5)}
$$

Из (3.5) видно, что вблизи плазменного резонанса, т.е. при $\Omega\approx1$, т.е. при
$\omega\approx\omega_p$ модуль нуля $|\eta_0(\alpha,\Omega,\varepsilon)|$
становится неограниченным при любых значениях безразмерного химического потенциала $\alpha$
в случае $\varepsilon\to 0$.

\begin{center}
  4. О СУЩЕСТВОВАНИИ ПЛАЗМЕННОЙ МОДЫ
\end{center}

Остановимся подробнее на вопросе существования плазменной моды, отвечающей конечному
комплексному нулю $\eta_0$ дисперсионной функции. Нуль $\eta_0$ является функцией
параметров исходной системы уравнений $\mu$, $\omega$ и $\nu$, или функцией параметров
$(\alpha,\Omega,\varepsilon)$.
Требуется найти область $D^+(\alpha)$, лежащую в
плоскости параметров $(\Omega,\varepsilon)$, такую, что если
$(\Omega,\varepsilon)\in D^+(\alpha)$, то число нулей $N$ дисперсионной функции
$\Lambda(z)$ равно двум: $N=2$. Через $D^-(\alpha)$ обозначим такую область на
плоскости параметров, что число нулей дисперсионной функции равно нулю: $N=0$.
Кривую, являющуюся границей этих областей, обозначим через $L=L(\alpha)$.

Множество физически значимых параметров $(\Omega,\varepsilon)$ заполняет
четверть-плоскость $\mathbb R^2_+ = \{(\Omega,\varepsilon):\Omega\geqslant0,
\varepsilon\geqslant0\}$. Случай $\Omega\geqslant0$ (или $\omega=0$) отвечает
внешнему стационарному электрическому полю, а случай $\varepsilon=0$ (или $\nu=0$)
отвечает случаю бесстолкновительной плазмы.

Возьмем контур $\mathrm\Gamma_{\rho}=C^+_{\rho}\cup C^-_{\rho}$, состоящий из двух
замкнутых полуокружностей $C^+_\rho$ и $C^-_\rho$  радиуса $R=1/\rho$,
лежащих соответственно в верхней и нижней полуплоскостях; $\rho$ - достаточно
малое действительное положительное число, $C^{\pm}_\rho=\{z=x+iy, |z|=1/\rho,
|x\pm i\rho|\leqslant 1/\rho\}$.
Число $R$ возьмем достаточно большим, чтобы нули дисперсионной функции
(если они существуют), лежали внутри области $D_\rho$, ограниченной контуром
$\mathrm\Gamma_\rho$. Заметим, что в пределе при $\rho\to0$ область  $D_\rho$
переходит в область $D_0$,
ограниченную контуром $\mathrm\Gamma_0=\lim\limits_{\rho\to0}\mathrm\Gamma_\rho$.
Эта область совпадает с комплексной плоскостью с разрезом вдоль действительной оси.

В силу принципа аргумента число \cite{Gahov} нулей $N$ дисперсионной функции в
области $D_\rho$ равно:

$$
N=\frac{1}{2\pi i}\oint\limits_{\mathrm\Gamma_\rho}d\ln\Lambda(z).
$$

Переходя к пределу в этом равенстве при $\rho\to0$ и учитывая, что дисперсионная
функция аналитична в окрестности бесконечно удаленной точки, получаем, что

$$
N=\frac{1}{2\pi i}\int_{-\infty}^{\infty}d\ln\Lambda^+(\tau)-
\frac{1}{2\pi i}\int_{-\infty}^{\infty}d\ln\Lambda^-(\tau).
$$

Во втором интеграле сделаем замену переменной $\tau\to-\tau$. Учитывая четность
дисперсионной функции, получаем, что
$$
N=\frac{1}{2\pi i}\int_{-\infty}^{\infty}d\ln\frac{\Lambda^+(\tau)}
{\Lambda^-(\tau)}.
$$

Разобьем этот интеграл на два - по полуосям $(-\infty,0)$ и $(0,+\infty)$.
В первом интеграле снова проведем указанную замену. Получаем, что
$$
N=\frac{1}{\pi i}\int_{0}^{\infty}d\ln\frac{\Lambda^+(\tau)}
{\Lambda^-(\tau)}=2\varkappa_{\mathbb R_+}(G).
\eqno {(4.1)}
$$
Здесь $\varkappa_{\mathbb R_+}(G)$ - индекс функции
$
G(\tau)={\Lambda^+(\tau)}/{\Lambda^-(\tau)},
$
вычисленный вдоль положительной действительной полуоси.

Таким образом, равенство (4.1) означает, что число нулей дисперсионной функции
$\Lambda(z)$ равно удвоенному индексу функции $G(\tau)$,
вычисленному вдоль положительной действительной полуоси.

Рассмотрим в комплексной плоскости кривую $\mathrm\Gamma_\alpha=\mathrm\Gamma(\alpha)$,
$$
\mathrm\Gamma(\alpha):z=G(\tau), \ 0\leqslant\tau\leqslant+\infty,
$$
Ясно, что $G(0)=1$, $\lim\limits_{\tau\to+\infty}G(\tau)=1$. Следовательно,
согласно (4.1), число нулей $N$ равно удвоенному числу оборотов кривой
$\mathrm\Gamma(\alpha)$ вокруг начала координат, т.е.
$$
N=2\varkappa(G), \qquad \varkappa(G)=\mathrm{ind}_{[0,+\infty]}G(\tau).
$$

Выделим у функции $G(\mu)$ действительную и мнимую части. Сначала представим функцию
$G(\mu)$ в виде:
$$
G(\mu)=\frac{\Omega^+(\mu}{\Omega^-(\mu)},
$$
где
$$
\Omega^\pm(\mu)=(w_0-1)\eta^2_1+(\eta^2_1-\mu^2)\lambda_0(\mu,\alpha)\pm
is(\mu,\alpha)(\eta^2_1-\mu^2),
$$
$$
s(\mu,\alpha)=\frac{\pi}{2s_0(\alpha)}\mu f_0(\mu,\alpha).
$$

Заметим, что
$$
w_0-1=-i\frac{\omega}{\nu}=-i\frac{\Omega}{\varepsilon}, \ \ \ \ \
\eta^2_1=\varepsilon r(\alpha)(\varepsilon-i\Omega),
$$
$$
(w_0-1)\eta^2_1=-\Omega^2r(\alpha)-i\varepsilon\Omega r(\alpha)=
-\Omega(\Omega+i\varepsilon)r(\alpha).
$$

Выделим действительную и мнимую части у функций $\Omega^\pm(\mu)$. Имеем:
$$
\Omega^\pm(\mu)=-P^\pm(\mu)-iQ^\pm(\mu),
$$
где
$$
P^\pm(\mu)=(1+\gamma)^2r(\alpha)+\lambda_0(\mu,\alpha)
(\mu^2-\varepsilon^2r(\alpha))\mp\varepsilon(1+\gamma)r(\alpha)s(\mu,\alpha),
$$
$$
Q^\pm(\mu)=\varepsilon(1+\gamma)r(\alpha)(1+\lambda_0(\mu,\alpha))\pm
(\mu^2-\varepsilon^2r(\alpha))s(\mu,\alpha).
$$

Теперь коэффициент $G(\mu)$ представим в виде:
$$
G(\mu)=\frac{P^+(\mu)+iQ^+(\mu)}{P^-(\mu)+iQ^-(\mu)}.
$$

Теперь легко выделить действительную и мнимую части функции $G(\mu)$:
$$
G(\mu,\alpha)=\frac{P^+P^-+Q^+Q^-}{(P^-)^2+(Q^-)^2}+i\frac{P^-Q^+-P^+Q^-}{(P^-)^2+(Q^-)^2},
$$
или, кратко,
$$
G(\mu)=G_1(\mu)+iG_2(\mu),
$$
где
$$
G_1(\mu)=\frac{g_1(\mu)}{g(\mu)}, \qquad
G_2(\mu)=\frac{g_2(\mu)}{g(\mu)}.
$$

Здесь
$$
g(\mu)=[P^-(\mu)]^2+[Q^-(\mu)]^2=$$$$=
[\Omega^2r(\alpha)+\lambda_0(\mu,\alpha)(\mu^2-\varepsilon^2r(\alpha))+
\varepsilon\Omega r(\alpha)s(\mu,\alpha)]^2+
$$
$$
+[\varepsilon\Omega(1+\lambda_0(\mu,\alpha))-s(\mu,\alpha)
(\mu^2-\varepsilon^2r(\alpha))]^2,
$$ \medskip
$$
g_1(\mu)=P^+(\mu)P^-(\mu)+Q^+(\mu)Q^-(\mu)=$$$$=
[\Omega^2r(\alpha)+\lambda_0(\mu,\alpha)(\mu^2-\varepsilon^2r(\alpha))]^2+
$$
$$
\hspace{3cm}+\varepsilon^2\Omega^2r^2(\alpha)[(1+\lambda_0(\mu,\alpha))^2-s^2(\mu,\alpha)]-
$$$$-
(\mu^2-\varepsilon^2r(\alpha))^2s^2(\mu,\alpha),
$$ \medskip
$$
g_2(\mu)=P^-(\mu)Q^+(\mu)-P^+(\mu)Q^-(\mu)=$$$$=
2s(\mu,\alpha)\big\{[\Omega2r(\alpha)+\lambda_0(\mu,\alpha)
(\mu^2-\varepsilon^2r(\alpha))]\times$$
$$\times(\mu^2-\varepsilon^2r(\alpha))+
\varepsilon^2\Omega^2r^2(\alpha)(1+\lambda_0(\mu,\alpha))\big\}.
$$

Выделим (см. рис. 3,4) на плоскости параметров $(\Omega,\varepsilon)$
однопараметрическое семейство кривую $L_{\alpha}=L(\alpha,\Omega,\varepsilon)$,
определяемую неявно заданными параметрическими уравнениями:
$$
L_\alpha=L_\alpha(\Omega,\varepsilon): \quad  g_1(\mu,\alpha,\Omega,\varepsilon)=0, \quad
g_2(\mu,\alpha,\Omega,\varepsilon)=0, \quad 0\leqslant\mu\leqslant+\infty,
$$
такую, что при переходе через эту кривую индекс функции $G(\mu)$ на
положительной полуоси меняется скачком.

Каждая кривая $L_\alpha$ разбивает плоскость параметров $(\Omega,\varepsilon)$ на
две области $D^+(\alpha)$ и $D^-(\alpha)$, такие, что при переходе точки
$(\Omega,\varepsilon)$ из одной области в другую индекс функции $G(\mu)$
на положительной полуоси меняется скачком.

Так же как и в \cite{Lat1998TMF} можно показать, что если
$(\Omega,\varepsilon)\in D^+(\alpha)$,
то $\varkappa_{[0,+\infty]}(G)=1$, а если $(\Omega,\varepsilon)\in D^-(\alpha)$,
то $\varkappa_{[0,+\infty]}(G)=0$. В первом случае (индекс равен единице) замкнутая
кривая $\mathrm\Gamma_\alpha$, введенная выше, один раз охватывает начало координат.
Во втором случае (индекс равен нулю) замкнутая кривая не охватывает начало координат.

Из равенства (4.1) вытекает, что число нулей дисперсионной функции равно двум
($N=2$), если $(\Omega,\varepsilon)\in D^+(\alpha)$, и дисперсионная функция нулей
не имеет, если $(\Omega,\varepsilon)\in D^-(\alpha)$.

Отметим, что в работе \cite{Lat1998TMF} разработан метод исследования граничного режима,
когда $(\Omega,\varepsilon)\in L_\alpha$.

Выведем явные параметрические уравнения кривой $L_\alpha$, разделяющей четверть-плоскость
параметров $(\Omega,\varepsilon)$ на две области $D^+(\alpha)$ и $D^-(\alpha)$.

Из уравнения $g_2(\mu,\alpha,\Omega,\varepsilon)=0$ находим:
$$
\Omega^2=-\frac{1}{r(\alpha)} \cdot
\frac{(\mu^2-\varepsilon^2r(\alpha))\lambda_0(\mu,\alpha)}{\mu^2+
\varepsilon^2r(\alpha)\lambda_0(\mu,\alpha)}.
\eqno {(4.2)}
$$

Возьмем уравнение $g_1(\mu,\alpha,\Omega,\varepsilon)=0$. Преобразуем это уравнение
с помощью равенства (4.2). Проделаем эту выкладку в общем виде. Из уравнения
$
g_1=P^-Q^+-P^+Q^-=0
$
находим, что $P^-=P^+({Q^-}/{Q^+})$. Далее находим, что
$$
 g_1\Big|_{g_2=0}=[P^-Q^+-P^+Q^-]\Bigg|_{P^-=\frac{Q^-}{Q^+}P^+}=
\frac{Q^-}{Q^+}\left[(P^+)^2+(P^-)^2\right].
$$

Ясно, что уравнение $g_1\big|_{g_2=0}=0$ эквивалетно уравнению
$Q^-(\mu)=0$. Из этого уравнения находим, что
$$
\Omega\varepsilon r(\alpha)[1+\lambda_0(\mu,\alpha)]=
(\mu^2-\varepsilon^2r(\alpha))s(\mu,\alpha).
$$
Возведем в квадрат обе части этого уравнения. В полученном уравнении выражение
$\Omega^2$ заменим согласно (4.2). После несложных преобразований получаем, что
$$
-\varepsilon^2r(\alpha)[1+\lambda_0(\mu,\alpha)]^2\frac{\lambda_0(\mu,\alpha)}
{\mu^2+\varepsilon^2r(\alpha)\lambda_0(\mu,\alpha)}=$$$$=s^2(\mu,\alpha)
(\mu^2-\varepsilon^2r(\alpha)),
$$
откуда находим следующую зависимость:
$$
\varepsilon^2=-\frac{1}{r(\alpha)} \cdot \frac{\mu^2s^2(\mu,\alpha)}
{\lambda_0(\mu,\alpha)[(1+\lambda_0(\mu,\alpha))^2+s^2(\mu,\alpha)]}.
\eqno {(4.3)}
$$

Теперь подставим квадрат параметра $\varepsilon^2$, определяемый равенством
(4.3), в (4.2). В результате получаем такую зависимость:
$$
\Omega^2=-\frac{1}{r(\alpha)} \cdot
\frac{\mu^2[\lambda_0(\mu,\alpha)(1+\lambda_0(\mu,\alpha))+
s^2(\mu,\alpha)]^2}{\lambda_0(\mu,\alpha)[(1+\lambda_0(\mu,\alpha))^2+s^2(\mu,\alpha)]}.
\eqno {(4.4)}
$$

Итак, нами найдены параметрические уравнения семейства кривых $L_\alpha$.
Из уравнений (4.3) и (4.4) находим эти уравнения:
$$
L_\alpha: \quad \Omega=\sqrt{L_1(\mu)}, \quad
\varepsilon=\sqrt{L_2(\mu)},\quad 0\leqslant\mu\leqslant+\infty.
\eqno {(4.5)}
$$

В (4.5) введены обозначения:
$$
L_1(\mu)=\frac{s_0(\alpha)}{s_2(\alpha)} \cdot
\frac{\mu^2[\lambda_0(\mu,\alpha)(1+\lambda_0(\mu,\alpha))+s^2(\mu,\alpha)]^2}
{[-\lambda_0(\mu,\alpha)][(1+\lambda_0(\mu,\alpha))^2+s^2(\mu,\alpha)]}
$$
и
$$
L_2(\mu)=\frac{s_0(\alpha)}{s_2(\alpha)} \cdot \frac{\mu^2s_2(\mu,\alpha)}
{[-\lambda_0(\mu,\alpha)][(1+\lambda_0(\mu,\alpha))^2+s^2(\mu,\alpha)]}.
$$

Итак, нами построена кривая $L_\alpha$, являющаяся границей областей $D^+(\alpha)$ и
$D^-(\alpha)$.
Напомним, что если $(\Omega,\varepsilon)\in D^+(\alpha)$, то
$$
\varkappa(G)=\mathrm{ind}_{[0,+\infty]}\frac{\Lambda^+(\mu)}{\Lambda^-(\mu)}=1.
$$
Это означает, что кривая $\mathrm\Gamma_\alpha$ один раз охватывает начало координат.
А если $(\Omega,\varepsilon)\in D^-(\alpha)$, то
$$
\varkappa(G)=\mathrm{ind}_{[0,+\infty]}\frac{\Lambda^+(\mu)}{\Lambda^-(\mu)}=0.
$$

Это означает, что кривая $\mathrm\Gamma_\alpha$ не охватывает начало координат.
Кривая $\mathrm\Gamma(\alpha)$ на комплексной плоскости $\mathbb C$ определяется уравнениями:
$$
\mathrm\Gamma_\alpha: \quad x=\mathrm{Re}\frac{\Lambda^+(\mu)}
{\Lambda^-(\mu)}, \quad y=\mathrm{Im}\frac{\Lambda^+(\mu)}
{\Lambda^-(\mu)}, \quad 0\leqslant\mu\leqslant+\infty.
$$

\begin{figure}
\begin{center}
\includegraphics[width=13.0cm, height=10cm]{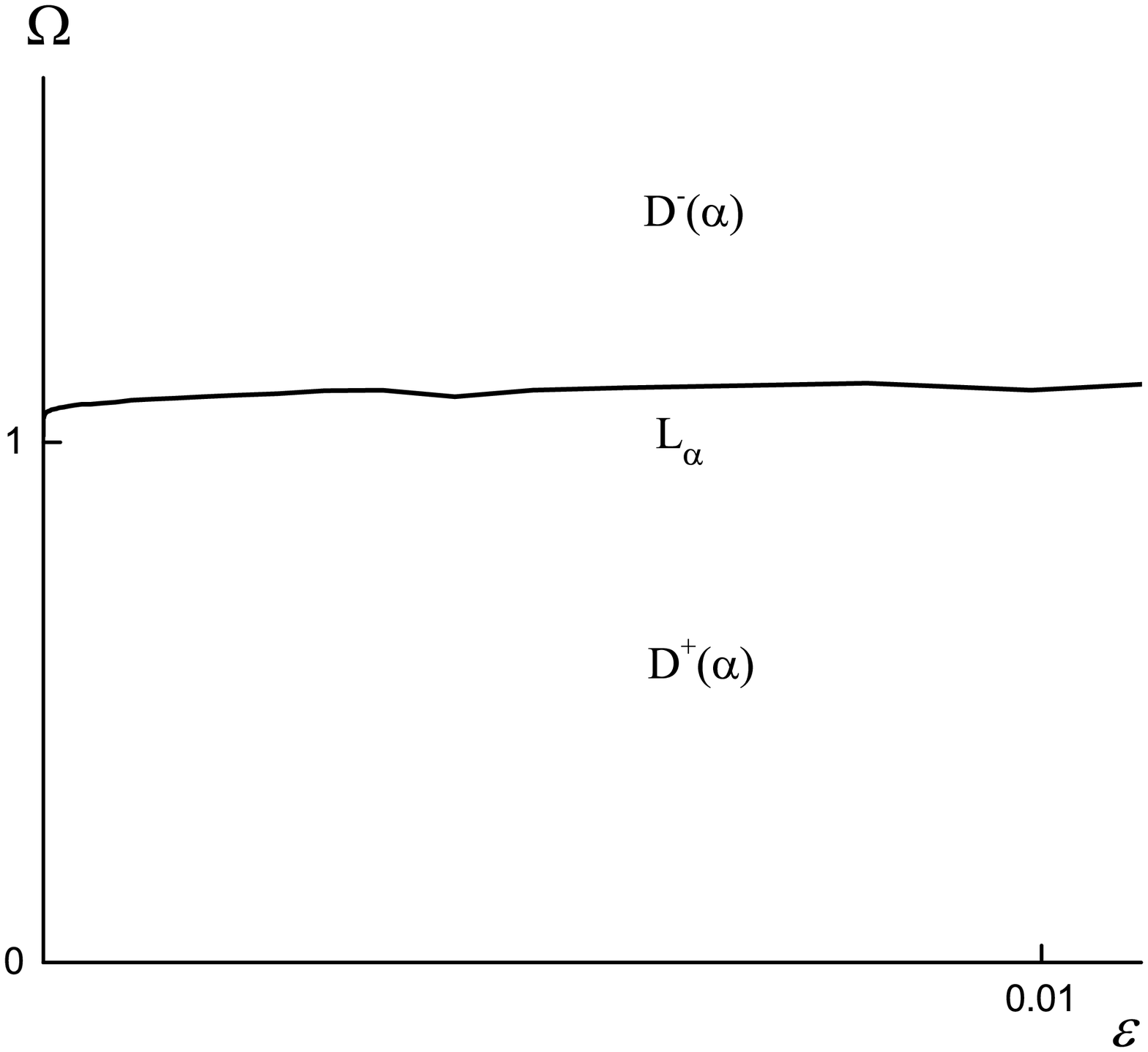}
\end{center}
\begin{center}
{Рис. 3. Кривая $L$ разделяет области $D^+$ и $D^-$. Случай $\alpha=-3$.}
\end{center}
\begin{center}
\includegraphics[width=13.0cm, height=10cm]{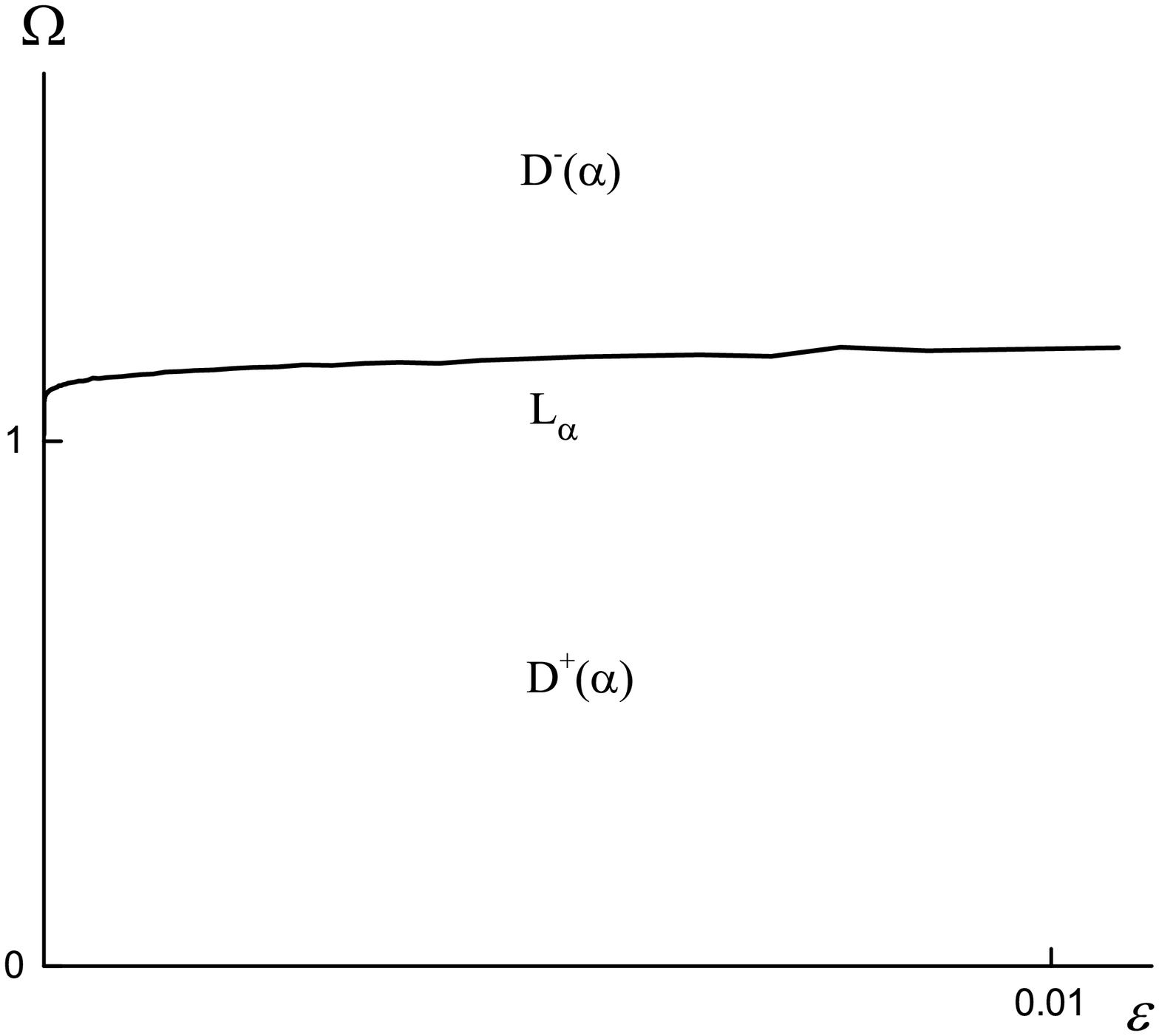}
\end{center}
\begin{center}
{Рис. 4. Кривая $L$ разделяет области $D^+$ и $D^-$. Случай $\alpha=3$.}
\end{center}
\end{figure}

\clearpage

Подчеркнем, что кривая $L_\alpha$ на плоскости параметров $(\Omega,\varepsilon)$
определяется параметрическими уравнениями:
$$
L(\alpha): \quad \mathrm{Re}\frac{\Lambda^+(\mu,\alpha,\Omega,\varepsilon)}
{\Lambda^-(\mu,\alpha,\Omega,\varepsilon)}=0, \quad
\mathrm{Im}\frac{\Lambda^+(\mu,\alpha,\Omega,\varepsilon)}
{\Lambda^-(\mu,\alpha,\Omega,\varepsilon)}=0, \quad 0\leqslant\mu\leqslant+\infty.
$$

Значение приведенного химического потенциала $\alpha$ при этом заполняет всю
замкнутую числовую прямую: $-\infty\leqslant\alpha\leqslant+\infty$.
При этом случай $\alpha$=$-\infty$ отвечает максвелловской плазме, а случай
$\alpha$=$+\infty$ отвечает полностью вырожденной плазме.

Сформулируем выводы в терминах плазменной (дебаевской, или дискретной) моды:
плазменная мода $H_{\eta_0}(x_1,\mu), e_{\eta_0}(x_1)$ существует (
или число нулей дисперсионной функции $\Lambda(z)$ равно двум, или индекс коэффициента
$G(\mu)=\Lambda^+(\mu)/\Lambda^-(\mu)$ равен единице на действительной полуоси), если
$(\Omega,\varepsilon)\in D^+(\alpha)$, и плазменная мода отсутствует (или число
нулей дисперсионной функции равно нулю, или индекс коэффициента равен нулю на
действительной полуоси), если $(\Omega,\varepsilon)\in D^-(\alpha)$.

\begin{center}
  5. ДИФФУЗНОЕ ОТРАЖЕНИЕ ЭЛЕКТРОНОВ ОТ ГРАНИЦЫ ПЛАЗМЫ
\end{center}

Будем решать задачу, состоящую из уравнений
(1.8) и (1.9) с граничными условиями (1.10)--(1.13). Решение задачи ищем в виде разложений
$$
H(x,\mu)=\dfrac{E_\infty}{w_0}\mu+\dfrac{E_0}{w_0}
\dfrac{\eta_0\mu-\eta_1^2}{\eta_0-\mu}
\exp\Big(-\dfrac{w_0x}{\eta_0}\Big)+$$$$+
\dfrac{1}{w_0}\int_{0}^{\infty}\exp\Big(-\dfrac{w_0x}{\eta}\Big)
F(\eta,\mu)\,E(\eta)\,d\eta,
\eqno{(5.1)}
$$
$$
e(x)=E_\infty+E_0\exp\Big(-\dfrac{w_0x}{\eta_0}\Big)+
\int_{0}^{\infty}\exp\Big(\dfrac{w_0x}{\eta}\Big)E(\eta)\,d\eta.
\eqno{(5.2)}
$$

Неизвестными в разложениях (5.1) и (5.2) являются коэффициенты
дискретного спектра $E_0,\; E_\infty$ и коэффициент непрерывного спектра $E(\eta)$,
причем $E_0=0$, если $(\Omega,\varepsilon)\in D^-(\alpha)$.

Подставляя собственные функции $F(\eta,\mu)$ в разложение (5.1), представим последнее в виде:
$$
H(x,\mu)=\dfrac{E_\infty}{w_0}\mu+\dfrac{E_0}{w_0}
\dfrac{\eta_0\mu-\eta_1^2}{\eta_0-\mu}
\exp\Big(-\dfrac{w_0x}{\eta_0}\Big)-
$$
$$
-\dfrac{1}{w_0}\int_{0}^{\infty}\exp\Big(-\dfrac{w_0x}{\eta}\Big)
\eta\,E(\eta)\,d\eta+$$$$+\dfrac{1}{w_0}\int_{0}^{\infty}
\exp\Big(-\dfrac{w_0x}{\eta}\Big)\dfrac{\eta^2-\eta_1^2}{\eta-\mu}
E(\eta)\,d\eta-
$$
$$
-\eta_1^2\exp\Big(-\dfrac{w_0x}{\mu}\Big)
\dfrac{\Lambda(\mu)}{\mu k(\mu,\alpha)}E(\mu)H(\mu), \quad -\infty<\mu<+\infty,\;x\geqslant 0.
\eqno{(5.3)}
$$

Подставим разложения (5.3) и (5.2) в соответствующие граничные
условия. Получаем следующую систему уравнений:
$$
w_0A=E_\infty\mu+E_0\dfrac{\eta_0\mu-\eta_1^2}{\eta_0-\mu}-
\int_{0}^{\infty}\eta E(\eta)\,d\eta+
$$
$$
+\int_{0}^{\infty}\dfrac{\eta^2-\eta_1^2}{\eta-\mu}E(\eta)\,d\eta-
w_0\eta_1^2\dfrac{\Lambda(\mu,\alpha)}{\mu k(\mu,\alpha)}E(\mu)H(\mu), \quad
0<\mu<+\infty,
\eqno{(5.4)}
$$
$$
E_\infty+E_0+\int_{0}^{\infty}E(\eta)\,d\eta=1.
\eqno{(5.5)}
$$

Рассмотрим условие непротекания (1.13).
Нам понадобится первый момент собственной функции непрерывного спектра. Имеем:
$$
\int_{-\infty}^{+\infty}\mu F(\eta,\mu)f_0(\mu,\alpha)\,d\mu=$$$$=
-\int_{-\infty}^{+\infty}
\dfrac{\mu(\mu\eta-\eta_1^2)}{\mu-\eta}f_0(\mu,\alpha)\,d\mu-
2\eta_1^2w_0s_0(\alpha)\Lambda(\eta)=
$$
$$
=2\eta_1^2s_0(\alpha)-2w_0\eta_1^2s_0(\alpha)[1-\Lambda(\eta,\alpha)]-
2\eta_1^2w_0s_0(\alpha)\Lambda(\eta)=
$$
$$
=2\eta_1^2s_0(\alpha)(1-w_0)=i2\eta_1^2s_0(\alpha)\dfrac{\Omega}{\varepsilon}.
\eqno{(5.6)}
$$\bigskip
Для собственной функции дискретного спектра
$F(\eta_0,\mu)$ имеет место аналогичное равенство:
$$
\int_{-\infty}^{\infty}\mu F(\eta_0,\mu)f_0(\mu,\alpha)\,d\mu=-\dfrac{1}{w_0}
\int_{-\infty}^{\infty}\mu\dfrac{\eta_0\mu-\eta_1^2}{\eta_0-\mu}f_0(\mu,\alpha)d\mu=
$$
$$
=2\eta_1^2(1-w_0)s_0(\alpha)=i2\eta_1^2s_0(\alpha)\dfrac{\Omega}{\varepsilon}.
\eqno{(5.7)}
$$
Заметим, что правые части соотношений (5.6) и (5.7) совпадают.

Представим условие непротекания электронов через границу в явном виде
$$
\int_{-\infty}^\infty \mu f_0(\mu,\alpha)\Bigg[E_\infty \mu+E_0\dfrac{\eta_0\mu-\eta_1^2}
{\eta_0-\mu}+\int_0^\infty F(\eta,\mu)E(\eta)d\eta\Bigg]d\mu=0,
$$
или, поменяв местами порядок интегрирования в третьем слагаемом, получаем, что
$$
E_\infty\int_{-\infty}^\infty \mu^2f_0(\mu,\alpha)d\mu+E_0\int_{-\infty}^\infty
\dfrac{\eta_0\mu-\eta_1^2}{\eta_0-\mu}\mu f_0(\mu,\alpha)d\mu+
$$
$$
+\int_0^\infty E(\eta)d\eta\int_{-\infty}^\infty \mu f_0(\mu,\alpha)F(\eta,\mu)d\mu=0.
\eqno{(5.8)}
$$
Первый интеграл из (5.8) равен:
$$
\int_{-\infty}^\infty\mu^2f_0(\mu,\alpha)d\mu=2s_2(\alpha).
$$
Второй интеграл из (5.8) вычисляется с помощью уравнения
$$
\Lambda(\eta_0)\equiv 1+\dfrac{\eta_0}{w_0\eta_1^2}\int_{-\infty}^\infty
\dfrac{\eta_1^2-\mu\eta_0}{\mu-\eta_0}k(\mu,\alpha)d\mu=0.
$$
С помощью этого уравнения получаем, что
$$
\int_{-\infty}^\infty\dfrac{\eta_0\mu-\eta_1^2}{\eta_0-\mu}\mu k(\mu,\alpha)d\mu=
\eta_1^2+\eta_0\int_{-\infty}^\infty\dfrac{\eta_1^2-\eta_0\mu}{\mu-\eta_0}k(\mu,\alpha)d\mu=
$$
$$
=\eta_1^2-w_0\eta_1^2=\eta_1^2(1-w_0)=\eta_1^2i\dfrac{\Omega}{\varepsilon}.
$$
Теперь ясно, что второй интеграл равен:
$$
\int_{-\infty}^\infty\dfrac{\eta_0\mu-\eta_1^2}{\eta_0-\mu}\mu f_0(\mu,\alpha)d\mu=
2s_0(\alpha)\eta_1^2i\dfrac{\Omega}{\varepsilon}.
$$
Внутренний интеграл из третьего слагаемого из (5.8) вычисляется по формуле (5.6). В
результате условие непротекания (5.8) упрощается и имеет вид:
$$
2s_2(\alpha)E_\infty+E_02s_0(\alpha)\eta_1^2i\dfrac{\Omega}{\varepsilon}+
2s_0(\alpha)\eta_1^2i\dfrac{\Omega}{\varepsilon}\int_0^\infty E(\eta)d\eta=0,
$$
или
$$
r(\alpha)E_\infty+i\eta_1^2\dfrac{\Omega}{\varepsilon}
\Big(E_0+\int_0^\infty E(\eta)d\eta\Big)=0.
\eqno{(5.9)}
$$

С помощью равенств (5.5) и (5.9) из условия непротекания получаем уравнение:
$$
r(\alpha)E_\infty+i\eta_1^2\dfrac{\Omega}{\varepsilon}(1-E_\infty)=0.
$$
Из этого уравнения находим амплитуду Друде:
$$
E_\infty=-\dfrac{i\eta_1^2(\Omega/\varepsilon)}{r(\alpha)-i\eta_1^2(\Omega/\varepsilon)}=
$$$$=
-\dfrac{i\Omega(\varepsilon-i\Omega)}{1-i\Omega(\varepsilon-i\Omega)}=-
\dfrac{\Omega(\Omega+i\varepsilon)}{1-\Omega(\Omega+i\varepsilon)}.
\eqno{(5.10)}
$$

Коэффициент дискретного спектра найден и определяется равенством (5.10).

Нетрудно проверить, что амплитуда Друде есть отношение двух значений дисперсионной
функции в точках $\eta=\eta_1$ и $\eta=\infty$:
$$
E_\infty=\dfrac{\Lambda(\eta_1)}{\Lambda(\infty)}=\dfrac{\Lambda_1}{\Lambda_\infty}.
$$
В самом деле, вычисляя это отношение, находим, что
$$
\dfrac{\Lambda_1}{\Lambda_\infty}=
\dfrac{1-{w_0}^{-1}}{1-{w_0}^{-1}+(w_0\varepsilon)^{-2}}=
-\dfrac{\Omega(\Omega+i\varepsilon)}{1-\Omega(\Omega+i\varepsilon)}.
$$

Формула (5.10) означает, что амплитуда Друде не зависит от граничных условий задачи, т.е.
от характера взаимодействия электронов с границей плазмы.

Из уравнений (5.5) и (5.10) находим:
$$
E_0+\int_{0}^{\infty} E(\eta)\,d\eta=1-E_\infty=1-\dfrac{\Lambda_1}{\Lambda_\infty}=
$$
$$
=\dfrac{1}{1+w_0\varepsilon^2(w_0-1)}=\dfrac{1}{w_0\varepsilon^2\Lambda_\infty}=
\dfrac{\Lambda_1}{\Lambda_\infty w_0\varepsilon^2(w_0-1)}.
\eqno{(5.11)}
$$

Вернемся к решению системы уравнений (5.4) и (5.11). Уравнение (5.4) является
сингулярным интегральным уравнением с ядром Коши, а уравнение (5.11) -- уравнением
Фредгольма.

Введем вспомогательную функцию
$$
N(z)=\int_{0}^{\infty}\dfrac{\eta^2-\eta_1^2}{\eta-z}E(\eta)\,d\eta,
\eqno{(5.12)}
$$
для которой разность граничных значений на числовой прямой равна:
$$
N^+(\mu)-N^-(\mu)=2\pi i (\mu^2-\eta_1^2)E(\mu),\quad 0<\mu<\infty.
\eqno{(5.13)}
$$

Преобразуем уравнение (5.4) к  краевому условию неоднородной
задачи Римана
$$
\Lambda^+(\mu)[N^+(\mu)+\varphi(\mu)]=
\Lambda^-(\mu)[N^-(\mu)+\varphi(\mu)],\quad 0<\mu<+\infty.
\eqno{(5.14)}
$$
Здесь
$$
\varphi(\mu)=-w_0A+E_\infty\mu+E_0\dfrac{\eta_0\mu-\eta_1^2}
{\eta_0-\mu}-\int_{0}^{\infty}\eta E(\eta)\,d\eta.
\eqno{(5.15)}
$$

Для решения неоднородной краевой задачи (5.14) сначала решим соответствующую
однородную краевую задачу.
Однородная краевая задача Римана здесь формулируется следующим
образом: требуется найти функцию $X(z)$, аналитическую везде в
комплексной плоскости, за исключением точек разреза $[0,+\infty]$,
имеющую в бесконечно удаленной точке порядок, равный индексу
функции $G(\mu)$ на отрезке $[0,1]$, и такую, что ее граничные
значения $X^{\pm}(\mu)$ сверху и снизу в точках интервала $(0,+\infty)$
связаны условием линейного сопряжения:
$$
X^+(\mu)=G(\mu)X^-(\mu), \qquad 0<\mu<+\infty,
\eqno{(5.16)}
$$
где, напомним,
$$
G(\mu)=\dfrac{\Lambda^+(\mu)}{\Lambda^-(\mu)}.
$$

Из условия (5.16) получаем счетное множество задач определения
аналитической функции по ее скачку на разрезе (при переходе через
разрез):
$$
\ln X^+(\mu)-\ln X^-(\mu)=$$$$=\ln G(\mu)+2\pi k i, \quad
0<\mu<+\infty,\quad k=0,\pm 1,\pm 2, \cdots.
\eqno{(5.17)}
$$

Для случая $\varkappa(G)=1$ возьмем $k=-1$, а для случая
$\varkappa(G)=0$ возьмем $k=0$. Тем самым в обеих случаях индекса
фиксируется такая регулярная непрерывная ветвь логарифма,
на которой выполняется условие:
$$
\ln G(+\infty)-2\pi i \varkappa(G)=0.
$$

В качестве решения задачи (5.17) возьмем интеграл типа Коши
$$
\ln X(z)=\dfrac{1}{2\pi i}\int_{0}^{\infty}
\dfrac{\ln G(\tau)-2\pi i \varkappa(G)}{\tau-z}\,d\tau.
\eqno{(5.18)}
$$

Правую часть этого равенства обозначим через $V(z)$:
$$
V(z)=\dfrac{1}{2\pi i}\int_{0}^{\infty}
\dfrac{\ln G(\tau)-2\pi i \varkappa(G)}{\tau-z}\,d\tau.
$$

Тогда
$$
X(z)=\exp V(z).
\eqno{(5.19)}
$$

Заметим, что решение (5.19) обращается в нуль в начале координат
при $\varkappa(G)=1$. Чтобы избежать этого недостатка
переопределим решение (5.19) следующим образом:
$$
X(z)=\dfrac{1}{z}\exp\Big(\dfrac{1}{2\pi i}\int_{0}^{\infty}
\dfrac{\ln G(\tau,\alpha)-2\pi i}{\tau-z}\,d\tau\Big).
\eqno{(5.20)}
$$

Итак, при $\varkappa(G)=0$ решение задачи (5.16) дается равенством
(5.19), а в случае $\varkappa(G)=1$ -- равенством (5.20). Оба
решения можно объединить в одно:
$$
X(z)=\dfrac{1}{z^\varkappa}\exp V(z),
$$
где функция $V(z)$ дается равенством (5.18).

С помощью решения соответствующей однородной краевой задачи
Римана сведем задачу (5.14) к решению задачи определения
аналитической функции по ее скачку, заданному на разрезе:
$$
X^+(\mu)[N^+(\mu)+\varphi(\mu)]=
X^-(\mu)[N^-(\mu)+\varphi(\mu)],\quad 0<\mu<+\infty.
\eqno{(5.21)}
$$
Здесь
$$
X(z)=\dfrac{1}{z}\exp V(z),
$$
где
$$
V(z)=\dfrac{1}{\pi}\int_{0}^{\infty}\dfrac{\zeta(\tau)\,d\tau}
{\tau-z}, \qquad \zeta(\tau)=\dfrac{1}{2i}\ln G(\tau)-\pi.
$$

Учитывая поведение функции $X(z)$ и функции $\varphi(z)$,
определенной равенством (5.15), получим общее решение задачи
(5.21):
$$
N(z)=w_0A-E_\infty z-E_0\dfrac{\eta_0 z-\eta_1^2}{\eta_0-z}+
\int_{0}^{\infty}\eta E(\eta)\,d\eta+$$$$
+\dfrac{1}{X(z)}\Big(C_0+\dfrac{C_{-1}}{z-\eta_0}\Big),
\eqno{(5.22)}
$$
где $C_0$ и $C_1$ -- произвольные постоянные.

Найдем асимптотику общего решения (5.22) при $z\to \infty$:
$$
N(z)=w_0A-E_\infty z+E_0\eta_0+\int_{0}^{\infty}\eta
E(\eta)\,d\eta+C_0z+
$$
$$
+C_{-1}-C_0V_1+o(1),\quad z\to \infty.
\eqno{(5.23)}
$$
Устраняя у решения (5.22) полюс в бесконечно удаленной точки с учетом
(5.23), получаем
$$
C_0=E_\infty.
\eqno{(5.24)}
$$

Из условия $N(\infty)=0$ и равенства (5.24) получаем уравнением Фредгольма:
$$
w_0A=-E_0\eta_0-\int_{0}^{\infty}\eta
E(\eta)\,d\eta-C_{-1}+E_\infty V_1,
\eqno{(5.25)}
$$
где
$$
V_1=-\dfrac{1}{\pi}\int_{0}^{\infty}\zeta(\tau)d\tau.
$$
Устраняя у решения (5.22) полюс в точке $\eta_0$, имеем:
$$
C_{-1}=-E_0(\eta_0^2-\eta_1^2)X(\eta_0).
\eqno{(5.26)}
$$

Осталось найти коэффициент дискретного спектра $E_0$, константу непротекания $A_+$ и
коэффициент непрерывного спектра $E(\eta)$.

Коэффициент непрерывного спектра $E(\eta)$ найдем, если подставим общее
решение (5.22) в формулу Сохоцкого (5.13), записанную для вспомогательной функции (5.12).
В результате находим:
$$
E(\eta)=\dfrac{1}{2\pi i(\eta^2-\eta_1^2)}\Big(C_0+
\dfrac{C_{-1}}{\eta-\eta_0}\Big)\Big(\dfrac{1}{X^+(\eta)}-
\dfrac{1}{X^-(\eta)}\Big).
\eqno{(5.27)}
$$

Для нахождения $E_0$ подставим (5.27) в условие (5.5). Имеем:
$$
E_\infty+E_0+\dfrac{1}{2\pi i}\int_{0}^{\infty}
\Big(\dfrac{1}{X^+(\eta)}-\dfrac{1}{X^-(\eta)}\Big)\Big(E_\infty-
$$$$-
\dfrac{E_0X(\eta_0)(\eta_0^2-\eta_1^2)}{\eta-\eta_0}\Big)\dfrac{d\eta}
{\eta^2-\eta_1^2}=1.
\eqno{(5.28)}
$$

Первый из интегралов (5.28) вычислим с помощью разложений на
элементарные дроби и интегрального представления.
Интегральное представление приведем без вывода:
$$
\dfrac{1}{X(z)}-z+V_1(\alpha)=$$$$=\dfrac{1}{2\pi i}\int_{0}^{\infty}
\Big(\dfrac{1}{X^+(\eta)}
-\dfrac{1}{X^-(\eta)}\Big)\dfrac{d\eta}{\eta-z},\quad
z\in \mathbb{C}\setminus [0,+\infty].
\eqno{(5.29)}
$$
Имеем:
$$
J_1=\dfrac{1}{2\pi i}\int_{0}^{\infty}
\Big(\dfrac{1}{X^+(\eta)}-\dfrac{1}{X^-(\eta)}\Big)\dfrac{d\eta}
{\eta^2-\eta_1^2}=\dfrac{1}{2\eta_1}\Big(J(\eta_1)-J(-\eta_1)\Big),
$$
где
$$
J(\pm\eta_1)=\dfrac{1}{2\pi i}\int_{0}^{\infty}
\Big(\dfrac{1}{X^+(\eta)}-\dfrac{1}{X^-(\eta)}\Big)\dfrac{d\eta}
{\eta \mp \eta_1}.
$$
Согласно (5.29) получаем:
$$
J(\pm \eta_1)=\dfrac{1}{X(\pm\eta_1)}\mp\eta_1+V_1.
$$
Следовательно, первый из интегралов из (5.28) вычислен:
$$
J_1=\dfrac{1}{2\eta_1}\Big[\dfrac{1}{X(\eta_1)}-\dfrac{1}{X(-\eta_1)}
-2\eta_1\Big]=$$$$=
-1-\dfrac{X(\eta_1)-X(-\eta_1)}{2\eta_1X(\eta_1)X(-\eta_1)}.
$$

Воспользуемся формулой факторизации дисперсионной функции
$$
\Lambda(z)=\Lambda_\infty (\eta_0^2-z^2)X(z)X(-z), \qquad z\in
\mathbb{C}\setminus [-\infty,+\infty].
$$

Подставляя в это равенство $z=\eta_1$, получим:
$$
\Lambda_1=\Lambda_\infty(\eta_0^2-\eta_1^2)X(\eta_1)X(-\eta_1),
$$
откуда
$$
X(\eta_1)X(-\eta_1)=\dfrac{E_\infty}{\eta_0^2-\eta_1^2}.
$$

Следовательно, с использованием обозначения
$$
\alpha^{\pm}=\dfrac{X(\eta_1)\pm X(-\eta_1)}{2},
$$
получаем
$$
J_1=-1-\dfrac{\alpha^-(\eta_0^2-\eta_1^2)}{\eta_1 E_\infty}.
$$

Второй интеграл из (5.28)
$$
J_2=\dfrac{1}{2\pi i}\int_{0}^{\infty}
\Big(\dfrac{1}{X^+(\eta)}-\dfrac{1}{X^-(\eta)}\Big)\dfrac{d\eta}
{(\eta^2-\eta_1^2)(\eta-\eta_0)}
$$
также можно вычислить разложением на элементарные дроби. Однако, проще
воспользоваться контурным интегрированием. Возьмем контур, изображенный на рис. 5.

\begin{figure}
\begin{center}
\includegraphics[width=13cm, height=13cm]{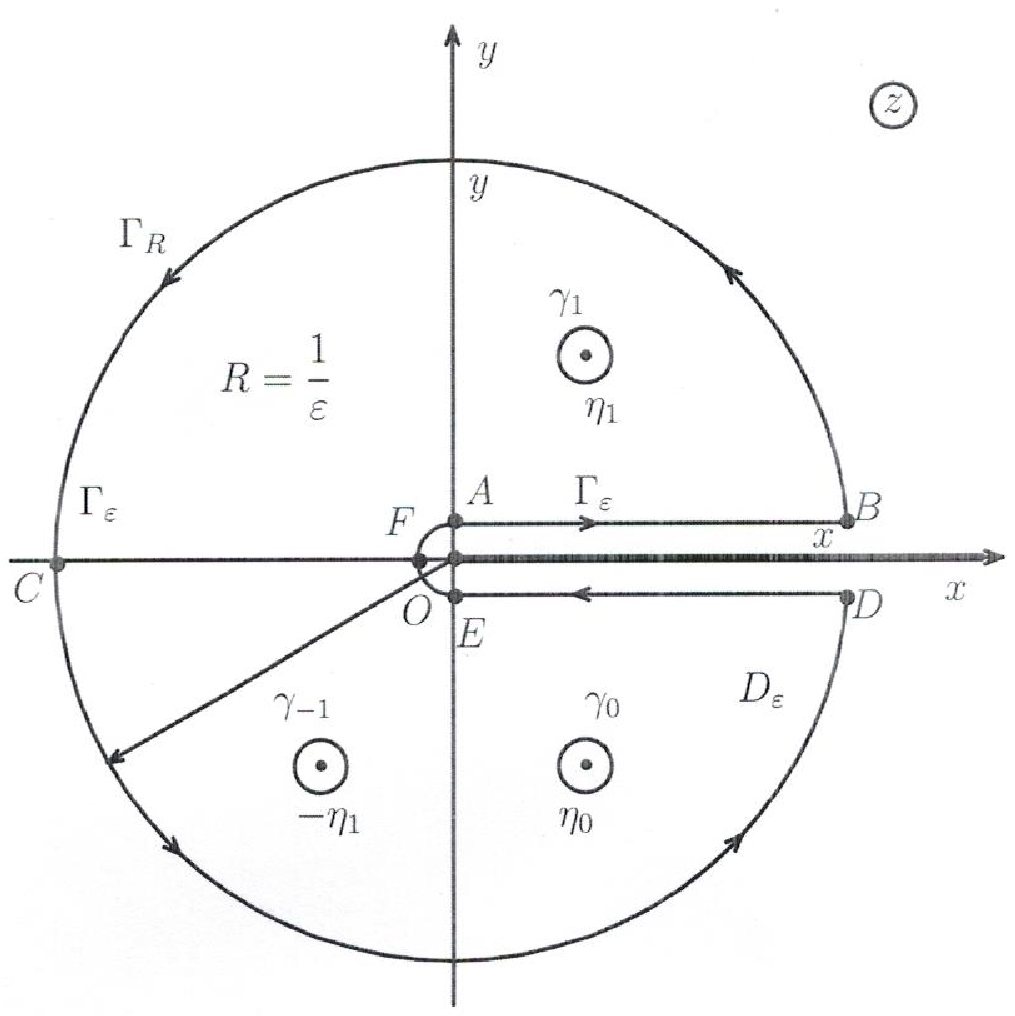}
\end{center}
\begin{center}
{Рис. 5. Контур интегрирования.}
\end{center}
\end{figure}

Рассуждая так же, как и выше, получаем:
$$
J_2=\Big[\Res_{\eta_1}+\Res_{-\eta_1}+\Res_{\eta_0}\Big]
\dfrac{1}{X(z)(z^2-\eta_1^2)(z-\eta_0)}=
$$
$$
=\dfrac{1}{2\eta_1}\Big[\dfrac{1}{X(\eta_1)(\eta_1-\eta_0)}+
\dfrac{1}{X(-\eta_1)(\eta_1+\eta_0)}\Big]+
\dfrac{1}{X(\eta_0)(\eta_0^2-\eta_1^2)}.
$$

С помощью (5.26) преобразуем этот интеграл:
$$
J_2=\dfrac{\eta_1[X(\eta_1)+X(-\eta_1)]-\eta_0[X(\eta_1)-X(-\eta_1)]}
{2\eta_1 X(\eta_1)X(-\eta_1)(\eta_1^2-\eta_0^2)}-\dfrac{E_0}{C_{-1}},
$$
или
$$
J_2=-\dfrac{\eta_1\alpha^+-\eta_0\alpha^-}{\eta_1E_\infty}+
\dfrac{1}{X(\eta_0)(\eta_0^2-\eta_1^2)}.
$$

Подставим значения интегралов $J_1$ и $J_2$ в (5.28). Получаем уравнение, из которого находим:
$$
E_0=\dfrac{E_\infty(\eta_1/(\eta_0^2-\eta_1^2)+\alpha^-)}
{X(\eta_0)(\eta_1\alpha^+-\eta_0\alpha^-)}.
\eqno{(5.30)}
$$

Из соотношений (5.30) и (5.26) следует, что
$$
C_{-1}=-\dfrac{E_\infty[\eta_1+\alpha^-(\eta_0^2-\eta_1^2)]}
{\eta_1\alpha^+-\eta_0\alpha^-}.
$$

Осталось найти константу непротекания $A$. Для этого требуется вычислить интеграл
$$
\int_{0}^{\infty}\eta\,E(\eta)\,d\eta=C_0
\int_{0}^{\infty}\Big(\dfrac{1}{X^+(\eta)}-\dfrac{1}{X^-(\eta)}\Big)
\dfrac{\eta\,d\eta}{\eta^2-\eta_1^2}+$$
$$
+C_{-1}\dfrac{1}{2\pi i}\int_{0}^{\infty}
\Big(\dfrac{1}{X^+(\eta)}-\dfrac{1}{X^-(\eta)}\Big)
\dfrac{\eta\,d\eta}{(\eta^2-\eta_1^2)(\eta-\eta_0)}.
\eqno{(5.31)}
$$

Первый интеграл вычисляется с помощью разложения на элементарные дроби.
Получаем, что
$$
\int_{0}^{\infty}\Big(\dfrac{1}{X^+(\eta)}-\dfrac{1}{X^-(\eta)}\Big)
\dfrac{\eta\,d\eta}{\eta^2-\eta_1^2}=\dfrac{J(\eta_1)+J(-\eta_1)}{2}=
$$
$$
=\dfrac{1}{2}\Big[\dfrac{1}{X(\eta_1)}+\dfrac{1}{X(-\eta_1)}+
2\eta_1\Big]=V_1+\dfrac{\alpha^+}{X(\eta_1)X(-\eta_1)}=$$$$=
V_1+\dfrac{\alpha^+}{E_\infty}(\eta_0^2-\eta_1^2).
$$

Второй интеграл проще вычислить с помощью вычетов:
$$
\dfrac{1}{2\pi i}\int_{0}^{\infty}
\Big(\dfrac{1}{X^+(\eta)}-\dfrac{1}{X^-(\eta)}\Big)\dfrac{\eta\,d\eta}
{(\eta^2-\eta_1^2)(\eta-\eta_0)}=
$$
$$=
\Big[\Res_{\infty}+\Res_{\eta_1}+\Res_{-\eta_1}+\Res_{\eta_0}\Big]
\dfrac{z}{(z^2-\eta_1^2)(z-\eta_0)}=$$$$
=-1+\dfrac{1}{2X(\eta_1)(\eta_1-\eta_0)}-\dfrac{1}{X(-\eta_1)
(\eta_1+\eta_0)}+$$$$+\dfrac{\eta_0}{X(\eta_0)(\eta_0^2-\eta_1^2)}=
-1-\dfrac{\eta_0\alpha^+-\eta_1\alpha^-}{E_\infty}+
\dfrac{\eta_0}{X(\eta_0)(\eta_0^2-\eta_1^2)}.
$$

Подставим значения найденных интегралов в (5.31). Имеем:
$$
\int_{0}^{\infty}\eta\,E(\eta)\,d\eta=
E_\infty V_1+\alpha^+(\eta_0^2-\eta_1^2)-$$$$
-C_{-1}-\dfrac{C_{-1}}{E_\infty}
(\eta_0\alpha^+-\eta_1\alpha^-)- \eta_0 E_0.
$$

Теперь подставим полученный интеграл в (5.25) находим, что
$$
w_0A=-\eta_1\dfrac{\eta_0\alpha^+-\eta_1\alpha^-+(\eta_0^2-\eta_1^2)
[(\alpha^+)^2-(\alpha^-)^2]}{\eta_1\alpha^+-\eta_0\alpha^-}.
$$

Используя обозначения величин $\alpha^{\pm}$ и формулу факторизации дисперсионной
функции, нетрудно видеть, что
$$
(\alpha^+)^2-(\alpha^-)^2=X(\eta_1)X(-\eta_1)=\dfrac{E_\infty}
{\eta_0^2-\eta_1^2}.
$$
Следовательно, константа непротекания определяется равенством:
$$
w_0A=-\eta_1\dfrac{E_\infty+\eta_0\alpha^+-\eta_1\alpha^-}
{\eta_1\alpha^+-\eta_0\alpha^-}.
\eqno{(5.32)}
$$

Теперь задача с диффузным отражением электронов решена полностью. Ее решение дается
разложениями (5.1) и (5.2). Коэффициенты этих разложений даются равенствами (5.10),
(5.27), (5.30) и (5.32).

\begin{center}
  ЗАКЛЮЧЕНИЕ
\end{center}

Итак, нами аналитически решена классическая задача о колебаниях электронной плазмы в
полупространстве, находящемся во внешнем переменном электрическом поле, и с диффузными
граничными условиями. В явном виде найдены функция распределения электронов и экранированное
электрическое поле внутри плазмы.
Функция распределения электронов и электрическое поле внутри плазмы представлены в
виде разложений по собственным функциям характеристической системы уравнений.
Коэффициенты разложений найдены с использованием граничных условий. Решение задачи
при этом сводится к решению сингулярного интегрального уравнения с ядром Коши.
Последнее преобразовано к неоднородной краевой задаче Римана. Для ее разрешения
сначала решается однородная краевая задача Римана (или задача факторизации коэффициента
краевой задачи). Коэффициенты разложения находятся из условий разрешимости краевой
задачи и существенного использования условия непротекания электронов через границу
плазмы.

Показано, что на плоскости параметров задачи $(\Omega, \varepsilon)$ существуют
области $D^(\alpha)$, разделенные привой $L_\alpha$, такие, что при
$(\Omega, \varepsilon) \in D^+(\alpha)$ мода Дебая
существует, а при $(\Omega, \varepsilon) \in D^-(\alpha)$ она исчезает.

Отметим, что помимо применяемого в работе метода кинетических уравнений используются
и другие подходы к изучению колебаний плазмы (см., например, \cite{Chizhonkov},
\cite{Morozov}). Развитый в работе метод может быть использован для решения
различных задач кинетической теории (см. \cite{Vedenyapin}).

\end{document}